\documentclass[11pt]{article}

\usepackage{graphicx}
\usepackage{hyperref}

\begin{document}

\baselineskip=17pt

%%%% *****************************************************************
%%%% *************    Text stat'i     ********************************
%%%% *****************************************************************
\newpage
\pagenumbering{arabic}

\begin{center}
\Large \bf Properties of particles in the ergosphere of black holes
\end{center}

\begin{center}
\bf
Andrey A. Grib${}^{1,2,\star}$\footnote{${}^{\phantom{\star} \star}$e-mail:\,
andrei\_grib@mail.ru}, \
Yuri V. Pavlov${}^{3,4,\star \star}$\footnote{${}^{\star \star}$e-mail:\,
yuri.pavlov@mail.ru}
\end{center}

\noindent
  ${}^{1}$\,A.\,Friedmann Laboratory for Theoretical Physics,
St.\,Petersburg, Russia\\
  ${}^{2}$\,Theoretical Physics and Astronomy Department,
The Herzen University, St.\,Petersburg, Russia\\
  ${}^{3}$\,Institute of Problems in Mechanical Engineering,
Russian Acad. Sci., St.\,Petersburg, Russia\\
  ${}^{4}$\,N.I. Lobachevsky Institute of Mathematics and Mechanics,
    Kazan Federal University, Kazan,\\ Russia

\begin{abstract}
    A new feature of the rotating black holes is the existence in their
ergosphere of trajectories of particles with negative and zero energies.
    Here we analyze general properties of such trajectories comparing
them with usual trajectories of particles with positive energy.
    A comparison with the situation in rotating coordinate frame in
Minkowski space-time is made.
    The possibility of the unbounded growth of the energy of two colliding
particles in the centre of mass frame in the ergosphere is analysed.
\end{abstract}

%%%% *****************************************************************
\section{Introduction}
\label{intro}

    As it is known special geodesics on which particles with negative and
zero energy (with nonzero angular momentum projection) can move are existing
in the ergosphere of rotating black holes~\cite{MTW}--\cite{NovikovFrolov}.
    Such particles surely are not observed in external to the ergosphere
region where only particles of positive relative to infinity energy can exist.
    However it occurs that one can find such geodesics in rotating
coordinates frame in Minkowski space-time.

    In this paper the systematic analysis of properties of such geodesics in
the ergosphere of black hole and comparison with case of rotating coordinates in
Minkowski space-time is made and some limitations on values of
the energy and angular momentum with any value of the energy are obtained.

    As it was shown by us earlier~\cite{GribPavlov2016}
the geodesics of particles with zero energy in the ergosphere of the black hole
as well as those for particles with negative energy~\cite{GribPavlovVert}
originate inside the gravitational radius, arrive to the ergosphere
and then go back inside the gravitational radius.
    Estimates for the angular velocities of particles with positive,
negative and zero energies are obtained.
    In the last section the problem of the unlimited growth of energy in
the centre of mass frame of two colliding particles for particles with
negative and zero energies is investigated.

    The system of units with gravitational constant and the light
velocity $G=c=1$ is used in the paper.

%%%% *****************************************************************
\section{Geodesics in Kerr's metric}
\label{sec2}

    The Kerr's metric of the rotating black hole~\cite{Kerr63}
in Boyer-Lindquist coordinates~\cite{BoyerLindquist67} is
    \begin{equation}
d s^2 = \frac{\rho^2 \Delta}{\Sigma^2}\, d t^2 -
\frac{\sin^2\! \theta}{\rho^2} \Sigma^2 \, ( d \varphi - \omega\, d t)^2
\label{Kerr}
- \frac{ \rho^2}{\Delta}\, d r^2 - \rho^2 d \theta^2 ,
\end{equation}
    where
    \begin{equation} \label{Delta}
\rho^2 = r^2 + a^2 \cos^2 \! \theta, \ \ \ \ \
\Delta = r^2 - 2 M r + a^2,
\end{equation}
    \begin{equation} \label{Sigma}
\Sigma^2 = (r^2 + a^2)^2 - a^2 \sin^2\! \theta\, \Delta , \ \ \ \
\omega = \frac{2 M r a}{\Sigma^2} ,
\end{equation}
    $M$ is the mass of the black hole, $ aM $ is angular momentum.
We suppose ${0 \le a \le M }$.
    The event horizon for the Kerr's black hole is given by
    \begin{equation}
r = r_H \equiv M + \sqrt{M^2 - a^2} .
\label{Hor}
\end{equation}
    The surface
    \begin{equation}
r = r_C \equiv M - \sqrt{M^2 - a^2} .
\label{HorC}
\end{equation}
    is the Cauchy horizon.
    The surface of the static limit is defined by
    \begin{equation}
r = r_1 \equiv M + \sqrt{M^2 - a^2 \cos^2 \theta} .
\label{Lst}
\end{equation}
    The region of space time between the static limit and the event horizon
is called ergosphere~\cite{MTW}, \cite{NovikovFrolov}.
    On the frontier of the ergosphere the function
    \begin{equation} \label{S}
S (r, \theta) = r^2 -2 M r + a^2 \cos^2 \! \theta
\end{equation}
    is zero, inside the ergosphere one has $ S (r, \theta) < 0$.
    Note that
    \begin{equation} \label{SigmaDr}
\Sigma^2 = (r^2 + a^2) \rho^2 + 2 r M a^2 \sin^2\! \theta > 0, \ \ {\rm if} \ \ r > 0.
\end{equation}

    Using the relation
    \begin{equation} \label{SSrho}
S\, \Sigma^2 + 4 M^2 r^2 a^2 \sin^2\! \theta = \rho^4 \Delta ,
\end{equation}
    one can write the equations of geodesics for the Kerr's metric~(\ref{Kerr})
(see~\cite{Chandrasekhar}, Sec.~62 or~\cite{NovikovFrolov}, Sec.~3.4.1)
as
    \begin{equation} \label{geodKerr1}
\rho^2 \frac{d t}{d \lambda } = \frac{1}{\Delta}
\left( \Sigma^2 E - 2 M r a J \right), \ \ \
\rho^2 \frac{d \varphi}{d \lambda } = \frac{1}{\Delta}
\left( 2 M r a E + \frac{S J}{\sin^2\! \theta} \right),
\end{equation}
    \begin{equation} \label{geodKerr3}
\rho^2 \frac{d r}{d \lambda} = \sigma_r \sqrt{R}, \ \ \ \ \
\rho^2 \frac{d \theta}{d \lambda} =\sigma_\theta \sqrt{\Theta},
\end{equation}
    \begin{equation} \label{geodR}
R = \Sigma^2 E^2 - \frac{S J^2}{\sin^2 \theta } - 4 M r a E J -
\Delta \left[ m^2 \rho^2 + \Theta \right],
\end{equation}
    \begin{equation} \label{geodTh}
\Theta = Q - \cos^2 \! \theta \left[ a^2 ( m^2 - E^2) +
\frac{J^2}{\sin^2 \! \theta} \right].
\end{equation}
    Here $E={\rm const}$ is the energy (relative to infinity) of the moving
particle,
$J$ is the conserved angular momentum projection on the rotation axis,
$m$ is the rest mass of the particle,
$\lambda $ --- the affine parameter along the geodesic.
    For the particle with $m \ne 0$ the parameter $\lambda = \tau /m$, where
$\tau$ is the proper time.
$Q$ is the Carter constant.
    $Q=0$ for the movement in the equatorial plane $(\theta = \pi/2)$.
    The constants $\sigma_r, \sigma_\theta = \pm 1$ define the direction
of movement on coordinates $r, \theta$.

%%%%%%%%%%%%%%%%%%%%%%%%%%%%%%%%%%%%%%%%%%%%%%%%%%
    From~(\ref{geodKerr3}) follows that the parameters characterizing any
geodesic must satisfy the conditions
    \begin{equation} \label{ThB0}
R \ge 0, \ \ \ \ \ \Theta \ge 0 .
\end{equation}
    For the geodesic being the trajectory of the test particle moving outside
the event horizon one must have the condition of movement ``forward in time''
    \begin{equation} \label{ThB0t}
d t / d \lambda > 0 .
\end{equation}
    The conditions~(\ref{ThB0}), (\ref{ThB0t}) lead to inequalities for
possible values of the energy $ E $ and angular momentum projection $J$ of
the test particle at the point with coordinates $(r, \theta)$ with fixed
value~$\Theta \ge 0$~\cite{GribPavlov2013}.

    Outside the ergosphere $ S(r, \theta) >0 $,
    \begin{equation} \label{EvErg}
E \ge \frac{1}{\rho^2} \sqrt{(m^2 \rho^2 + \Theta) S},
\ \  \ J \in \left[ J_{-} (r,\theta), \ J_{+} (r,\theta) \right],
\end{equation}
    \begin{equation}
J_{\pm} (r,\theta) = \frac{\sin \theta}{S} \left[ - 2 r M a E \sin \theta \pm
\sqrt{ \Delta \left( \rho^4 E^2 - (m^2 \rho^2 + \Theta) S \right)} \right].
\label{Jpm}
\end{equation}

    On the frontier of the ergosphere (for $\theta \ne 0, \pi$)
    \begin{equation} \label{JgErg}
r = r_1(\theta) \  \Rightarrow \  E \ge 0, \ \
J \le E \left[ \frac{M r_1(\theta) }{a} + a \sin^2 \! \theta \left(
1 - \frac{m^2}{2 E^2} - \frac{\Theta}{4 M r_1(\theta) E^2} \right) \right].
\end{equation}
    The value $E=0$ is possible on the frontier of the ergosphere when $m=0$,
$\Theta=0$.
    In this case one can have any value of $J <0$.

    Inside the ergosphere $ r_H < r < r_1(\theta) $, $S < 0$
    \begin{equation} \label{lHmdd}
J \le \frac{\sin \theta}{- S} \left[ 2 r M a E \sin \theta -
\sqrt{ \Delta \left( \rho^4 E^2 - (m^2 \rho^2 + \Theta) S \right)} \right].
\end{equation}
    and the energy of the particle as it is known can be as positive
as negative.

    As it is seen from~(\ref{JgErg}), (\ref{lHmdd}) on the frontier and inside
the ergosphere the angular momentum projection of particles moving along
geodesics can be negative and it can be any large in absolute value number
for the fixed value of the energy.
    This property first found in~\cite{GribPavlov2013,GribPavlov2012} for
the Kerr's metric (as it was shown later in~\cite{Zaslavskii13c}) is valid
in the ergosphere of any black hole with the axially symmetric metric.

    Note that in the vicinity of horizon from~(\ref{lHmdd}) one has
    \begin{equation} \label{JH}
J(r) \le J_H = \frac{2 r_H M E}{a}, \ \ \ r \to r_H.
\end{equation}

    If the the values of $J$ and $\Theta \ge 0$ are given then
from~(\ref{ThB0}), (\ref{ThB0t}) one has at any point outside the horizon
    \begin{equation}
E \ge \frac{1}{\Sigma^2} \left[ 2 M r a J + \sqrt{ \Delta \left(
\frac{ \rho^4 J^2 }{\sin^2\! \theta } + \left( m^2 \rho^2 + \Theta \right)
\Sigma^2 \right) } \right].
\label{EVnHS}
\end{equation}
    The lower frontier $E$ corresponds to $R=0$.
    As it is seen from~(\ref{EVnHS}) negative energies can be
only in case of the negative value of $J$ of the angular momentum of
the particle and for $r$ such as $2 M r a |\sin \theta| > \rho^2 \sqrt{\Delta}$,
i.e., in accordance with~(\ref{SSrho}), in the ergosphere.

%%%% *****************************************************************
\section{General limitations on angular and radial velocities for
the metric with rotation}
\label{seOmeg}

    Let us find limitations on the angular velocity of any particle in the metric
    \begin{equation}\label{gik}
d s^2 = g_{00}\, d t^2 + 2 g_{0 \varphi}\, d t d \varphi + g_{\varphi \varphi}\,
d \varphi^2 + g_{rr}\, d r^2 + g_{\theta \theta}\, d \theta^2
\end{equation}
    with $ g_{\varphi \varphi} < 0 $, $g_{rr} < 0$, $ g_{\theta \theta} < 0$
from the condition $ds^2 \ge 0$.
    Then
    \begin{equation}
g_{00}\, d t^2 + 2 g_{0 \varphi}\, d t\, d \varphi +
g_{\varphi \varphi}\, d \varphi^2 \ge 0
\label{omN1}
\end{equation}
    and the angular velocity $\Omega = d \varphi / d t $
for any particle is in the limits~\cite{MTW}
    \begin{equation}
\Omega_1 \le \Omega \le \Omega_2, \ \ \ \
\Omega_{1,2} = \frac{g_{0 \varphi} \mp \sqrt{ g_{0 \varphi}^{\,2} -
g_{00} g_{\varphi \varphi} }}{- g_{\varphi \varphi} }.
\label{omN2}
\end{equation}
    On the frontier of the ergosphere $g_{00} =0 $ and $\Omega_1 =0$.
    Inside the ergosphere $g_{00} < 0 $ and $\Omega_{1,2} > 0$
and all particles rotate in the direction of rotation of the black
hole~\cite{MTW}--\cite{NovikovFrolov}.

    Putting the components of the metric~(\ref{Kerr}) into~(\ref{omN2})
one obtains the limiting values of angular velocities for the rotating
Kerr black hole
    \begin{equation}
\Omega_{1,2} = \frac{2 M r a \sin \theta \mp \rho^2 \sqrt{\Delta}}
{\sin \theta \, \Sigma^2} = \omega \mp \frac{ \rho^2 \sqrt{\Delta}}
{\sin \theta \, \Sigma^2}.
\label{OmK1}
\end{equation}
    Approaching the event horizon one has
    \begin{equation}
\lim \limits_{\ \  r \to r_H} \Omega_1 (r) = \! \!
\lim \limits_{\ \  r \to r_H} \Omega_2 (r) = \omega_{\rm Bh}
= \frac{a}{2 M r_H}.
\label{omN3}
\end{equation}
    The value $\omega_{\rm Bh}$ is called angular velocity of rotation
of the black hole.

    Note that due to nonradial movement of massless particles in ergosphere
their radial velocity can have different values (zero for circular orbits).

    Let us find limitations on possible values of the radial velocities of
particles in the metric~(\ref{gik}) from the condition
    \begin{equation}
g_{00} + 2 g_{0 \varphi} \frac{d \varphi}{d t} +
g_{\varphi \varphi}\left( \frac{d \varphi}{d t} \right)^2  +
g_{rr} \left( \frac{d r}{d t} \right)^2 \ge 0.
\label{radv}
\end{equation}
    Taking into account that the sum of first three terms in~(\ref{radv})
has the maximal value at
    \begin{equation}
\frac{d \varphi }{ d t} = \frac{\Omega_2 - \Omega_1}{2},
\label{OmMar}
\end{equation}
    one obtains
    \begin{equation}
\left( \frac{d r }{ d t} \right)^2 \le \frac{g_{\varphi \varphi}}{g_{rr}}
\frac{(\Omega_2 - \Omega_1)^2}{4} = \frac{g_{0 \varphi}^{\,2} -
g_{00} g_{\varphi \varphi}}{g_{rr} g_{\varphi \varphi}}.
\label{RadMax}
\end{equation}

    For the Kerr's metric~(\ref{Kerr}) outside the event horizon $r > r_H$
one obtains
    \begin{equation}
\left( \frac{d r }{ d t} \right)^2 \le \frac{\Delta^2}{\Sigma^2} .
\label{RadMaKerr}
\end{equation}
    Close to the event horizon $\Delta (r) \to 0$
and so for any moving particle one has
    \begin{equation}
\frac{d r }{ d t} \to 0, \ \ {\rm if } \ \ r \to r_H.
\label{RadrH}
\end{equation}
    For particles moving on geodesics one obtains from
Eqs.~(\ref{geodKerr1})--(\ref{geodR})
    \begin{equation}
\left( \frac{d r }{ d t} \right)^2 =
\frac{ \Delta^2 }{ \Sigma^2 } \left[ 1 - \frac{ \Delta }{ \left( \Sigma^2 E -
2 M r a J \right)^2 } \left( \frac{\rho^4 J^2}{\sin^2 \! \theta } +
\Sigma^2 \left( m^2 \rho^2 + \Theta \right) \right) \right] .
\label{RadKerr}
\end{equation}
    So the maximal radial velocity $ |v_r|_{\rm max} = \Delta / \Sigma$
is obtained by massless particles with zero projection of the angular
momentum moving with fixed value of the angle $\theta$ ($\Theta=0$):
    \begin{equation}
\left| \frac{d r }{ d t} \right|_{\rm max} = \frac{\Delta}{ \Sigma}, \ \
{\rm if } \ \ m=0, \ \ J=0, \ \ Q = - a^2 E^2 \cos^2\! \theta.
\label{RadvMa}
\end{equation}

%%%% *****************************************************************
\section{Particles with zero energy in the Kerr's metric}
\label{secNulEn}

    In literature on black holes there is practically no discussion of
properties of particles with zero relative to infinity energy.
    Here we shall discuss geodesics for such particles.

    Values $E=0$ are possible on the frontier and inside the ergosphere.
    From the inequality $\Theta \ge 0$ and Eq.~(\ref{geodTh}) one obtains
    \begin{equation}
E =0 \ \ \Rightarrow \ \ Q \ge m^2 a^2 \cos^2 \theta + J^2 \cot^2 \theta \ge 0,
\label{E0Q}
\end{equation}
    i.e. for test particles with zero relative to infinity energy
the Carter constant is non negative.

    From the inequality~(\ref{lHmdd}) one obtains for particles with zero
energy inside ergosphere
    \begin{equation}
E =0 \ \ \ \Rightarrow \ \ \ J \le - \sin \theta \sqrt{\frac{\Delta
\left(m^2 \rho^2 + \Theta \right) }{-S} },
\label{E0Jtoch}
\end{equation}
    i.e. the angular momentum projection of particles with negative
energy is negative.

    It was shown by us earlier~\cite{GribPavlov2016} that geodesics of
particles with zero energy as well as as for particles with
negative~\cite{GribPavlovVert} energy originate and terminate on~$r_H$.
    One also obtains
    \begin{equation}
E =0 \ \ \Rightarrow \ \ R = - \Delta ( m^2 \rho^2 + \Theta ) -
\frac{J^2}{\sin^2 \theta } S(r, \theta) .
\label{E0R}
\end{equation}
    So the upper point of the trajectory of massive particles with zero
energy is located inside the ergosphere $( r < r_1(\theta ))$.
    Movement along the coordinate $\theta$  can be prolonged up to the point
with $\Theta (\theta) $ equal to zero.
    So in case $m=0$ from~(\ref{E0R}) one obtains that
 the upper point of the trajectory of massless particles with
zero energy is located on the frontier of the ergosphere  $ r_1(\theta )$.

    The radial velocity of the particle with zero energy is
    \begin{equation}
E =0 \ \ \Rightarrow \ \ \frac{d r}{d t} = \frac{ \sigma_r \Delta }{2 M r a}
\sqrt{ \frac{-S}{\sin^2\! \theta} - \Delta \frac{ m^2 \rho^2 + \Theta}{J^2} }
\label{rtE0}
\end{equation}
    and in general depends on the angular momentum.
    Indeed in case $m^2 \rho^2 + \Theta >0$ at the point with given $r$
in the ergosphere the radial velocity is the larger in absolute value
the lager is  $|J|$ and can vary from zero for the maximal value
$J= - \sin\! \theta \sqrt {\Delta (m^2 \rho^2 + \Theta )  / (- S) }$ to
    \begin{equation}
\frac{d r}{d t} = \frac{\sigma_r \Delta \sqrt{-S} }{2 M r a \sin \theta},
\ \ \ \ J \to - \infty .
\label{rtE0li}
\end{equation}

    From equations for geodesics it follows that
    \begin{equation}
E=0 \ \ \Rightarrow \ \
\frac{d \varphi}{d t}  = \frac{- S(r)}{2 r M a \sin^2 \! \theta}
= \frac{r^2 - 2 r M + a^2 \cos^2\! \theta}{- 2 r M a \sin^2 \! \theta}.
\label{omE0}
\end{equation}
    So {\it the angular velocity of any particle with zero energy
does not depend on the value of mass and angular momentum of the particle
and is defined by the formula~(\ref{omE0}).}

    If the particle with zero energy approaches the frontier of the ergosphere
(then it is necessary that $m=0$) one has $S(r)=0$ and
    \begin{equation}
r\to r_1(\theta) \ \ \Rightarrow \ \ \frac{d \varphi}{d t} \to 0.
\label{omE0lim}
\end{equation}
%%%%%%%%%%%%%%%%%%%%%%%%%%%%%%%%%%%%%%%%%%%%%%%%%%%%%%%%%%%%%%%%%%%%%%
    Inside the ergosphere the angular velocity of the particle with zero energy
as for any particle is positive.

    If
    \begin{equation}
E=0, \ \ Q = \cos^2 \! \theta \left[ a^2  m^2 + \frac{J^2}{\sin^2 \! \theta}
\right],
\label{EQtr0}
\end{equation}
    then $\Theta =0$ and from~(\ref{geodKerr3})
it follows that movement along geodesic occurs for constant~$\theta$.
    The form of trajectory of such photons
does not depend on the value of angular momentum projection as it is seen
from~(\ref{geodKerr1}), (\ref{geodKerr3}).
    The equation of this trajectory can be expressed in elementary functions
    \begin{eqnarray}
\varphi(r) - \varphi(r_1(\theta)) &=& \frac{- \sigma_r}{\sin \theta}
\left[ \arcsin \frac{ r - M}{\sqrt{M^2 - a^2 \cos^2 \theta}} - \frac{\pi}{2}
\right. + \nonumber \\
&&
+\left. \frac{a \sin \theta}{\sqrt{ M^2 - a^2 }}
\tanh^{-1} \frac{\sqrt{(M^2 -a^2) (2 r M - r^2 - a^2 \cos^2 \theta) }
}{a ( r - M) \sin \theta } \right], \ \ \ a < M,
\label{prea9}
\end{eqnarray}
    and is most simple for $a=M$:
    \begin{equation}
\varphi(r) - \varphi(r_1(\theta)) = \frac{- \sigma_r}{\sin \theta}
\left[ \arcsin \frac{r-M}{\sin \theta} +
\frac{\sqrt{2 r M - r^2 - M^2 \cos^2 \theta}}{r-M} - \frac{\pi}{2}\right].
\label{pre0}
\end{equation}
    The corresponding coordinate time of photon movement is
    \begin{eqnarray}
t (r) - t (r_1(\theta)) &=& \frac{- \sigma_r M }{\sqrt{ M^2 - a^2 }}
\left[ r_H \ln \frac{ M r_H - r \sqrt{ M^2 - a^2 } - a^2 \cos^2 \theta +
a \sin \theta \sqrt{-S(r, \theta)} }{M (r - r_H) } \right. -
\nonumber \\
&&
- \left. r_C \ln \frac{
M r_C + r \sqrt{ M^2 - a^2 } - a^2 \cos^2 \theta +
a \sin \theta \sqrt{-S(r, \theta)} }{M (r - r_C) } \right], \ \ \ a < M.
\label{pret9}
\end{eqnarray}
    When approaching the event horizon $ r \to r_H$ the time of movement
diverges logarithmically.
    Remind that the time of approaching the event horizon by the light ray in
Schwarzschild metric also diverges logarithmically~\cite{LL_II}.

    For the extremal rotating black hole
    \begin{equation}
t (r) - t (r_1(\theta)) = - \sigma_r 2 M \left[  \frac{ \sqrt{-S(r, \theta)}}{
(r - M) \sin \theta} - \ln \frac{ r - M }{ \sqrt{-S(r, \theta)} +
M \sin \theta } \right] .
\label{pret1A}
\end{equation}
    The time of approaching the event horizon $ r \to r_H$ here is divergent
as $\sim 1/ (r - r_H)$.

\begin{figure}[ht]
\centering
\includegraphics[width=50mm]{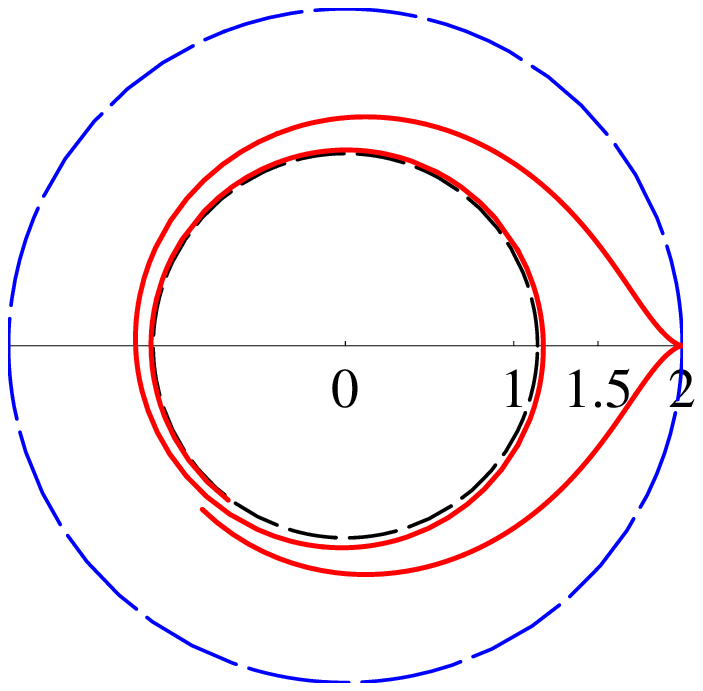} \ \ \ \
\includegraphics[width=75mm]{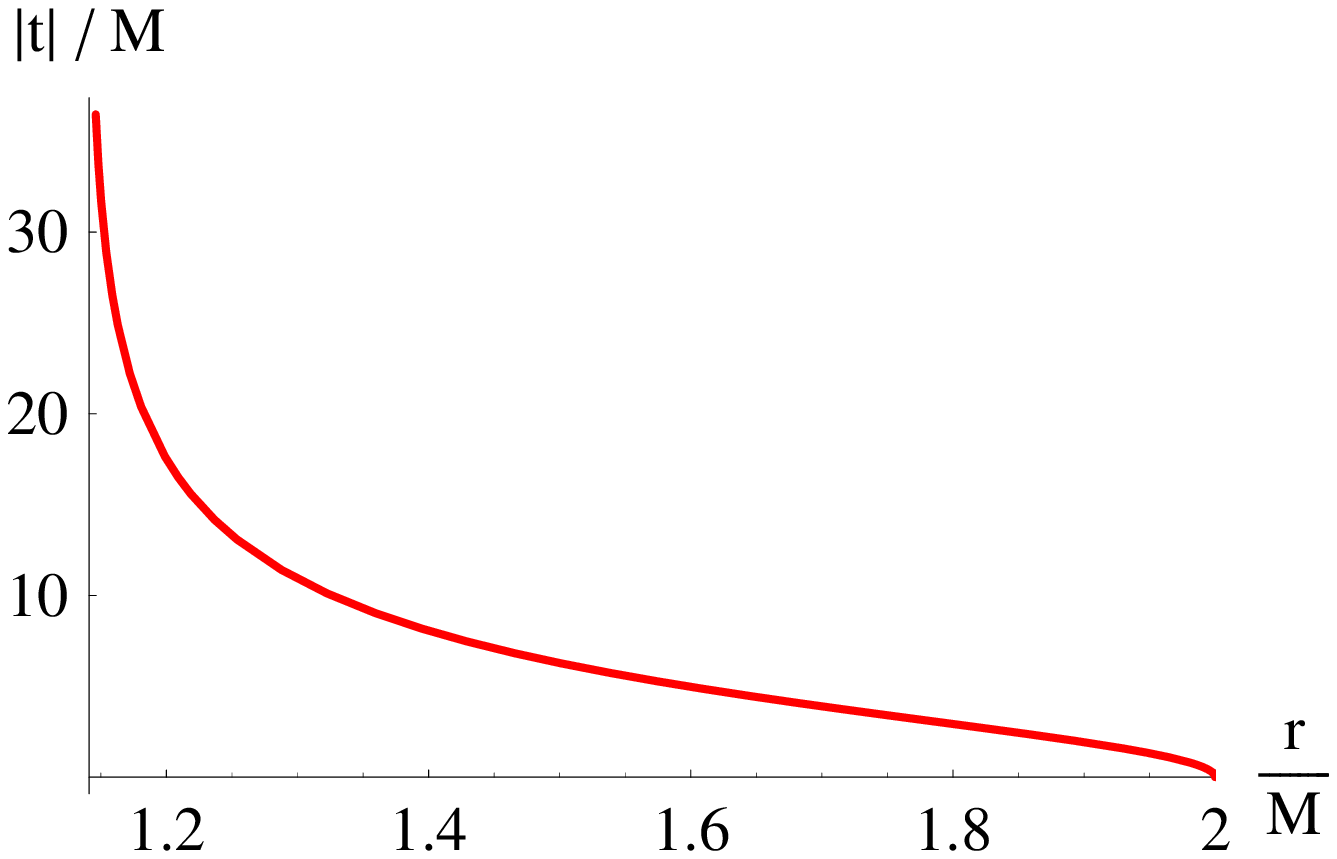}
\caption{The trajectories (at the left) of photons in the ergosphere
of the black hole
with $a=0.99 M$ and $E=0$ and time of movement (at the right).}
\label{Fig}
\end{figure}

    Part of the trajectory of the photon moving in equatorial
plane on geodesic with zero energy and the time of movement are
shown on Fig.~\ref{Fig}.
%%%%%%%%%%%%%%%%%%%%%%%%%%%%%%%%%%%%%%%%%%%%%%%%%
    The radial coordinate is used $r/M$.
    This trajectory as well as trajectories of any particles intersecting
the event horizon of the black hole rotates around the horizon infinitely
many times in Boyer-Lindquist coordinate frame.
    Really from the equations of geodesics~(\ref{geodKerr1}),
(\ref{geodKerr3}) it follows that in the vicinity of the horizon
    \begin{equation}
\frac{d \varphi}{d r} \sim \frac{1}{\Delta} = \frac{1}{(r-r_H)(r-r_C)}.
\label{PhirH}
\end{equation}
     Let us investigate the problem what simple properties of movement define
the difference of particles with negative, zero and positive relative to
the infinity energy.

%%%% *****************************************************************
\section{Angular velocity of particles with zero and negative energies}
\label{secNuOt}

    From equations of geodesics one obtains for the angular velocity of
freely moving particles
    \begin{equation}
\frac{d \varphi }{d t} = \frac{2 M r a E +
\frac{S J}{\sin^2\! \theta} }{\displaystyle
\Sigma^{\mathstrut 2} E - 2 M r a J } =
\omega + \frac{ \Delta \rho^4 J }{\sin^2\! \theta \Sigma^2
( \Sigma^2 E - 2 M r a J )} .
\label{OmK2}
\end{equation}

    From the limitation~(\ref{lHmdd}) one obtains for particles with
negative energy
    \begin{equation}
\frac{J}{E M} \ge \frac{\sin \theta}{- S(r)} \left[
2 r a \sin \theta + \frac{1}{M} \sqrt{ \Delta \left[
\rho^4 - \left( \frac{m^2}{E^2} \rho^2 + \frac{\Theta}{M^2 E^2} \right)
S(r) \right] } \right],
\label{OmK3}
\end{equation}
    and for particles with positive energy in the ergosphere
    \begin{equation}
\frac{J}{E M} \le \frac{\sin \theta}{- S(r)} \left[
2 r a \sin \theta - \frac{1}{M} \sqrt{ \Delta \left[
\rho^4 - \left( \frac{m^2}{E^2} \rho^2 + \frac{\Theta}{M^2 E^2} \right)
S(r) \right] } \right].
\label{OmK4}
\end{equation}

\begin{figure}[ht]
\centering
\includegraphics[width=60mm]{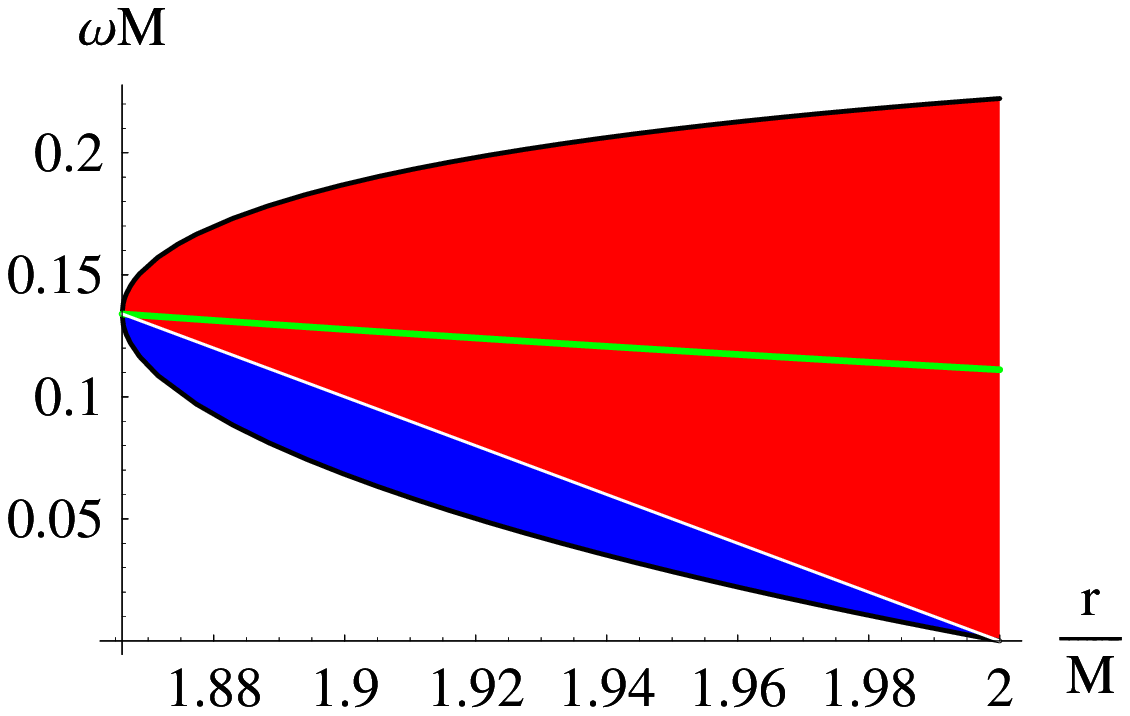} \ \ \ \
\includegraphics[width=60mm]{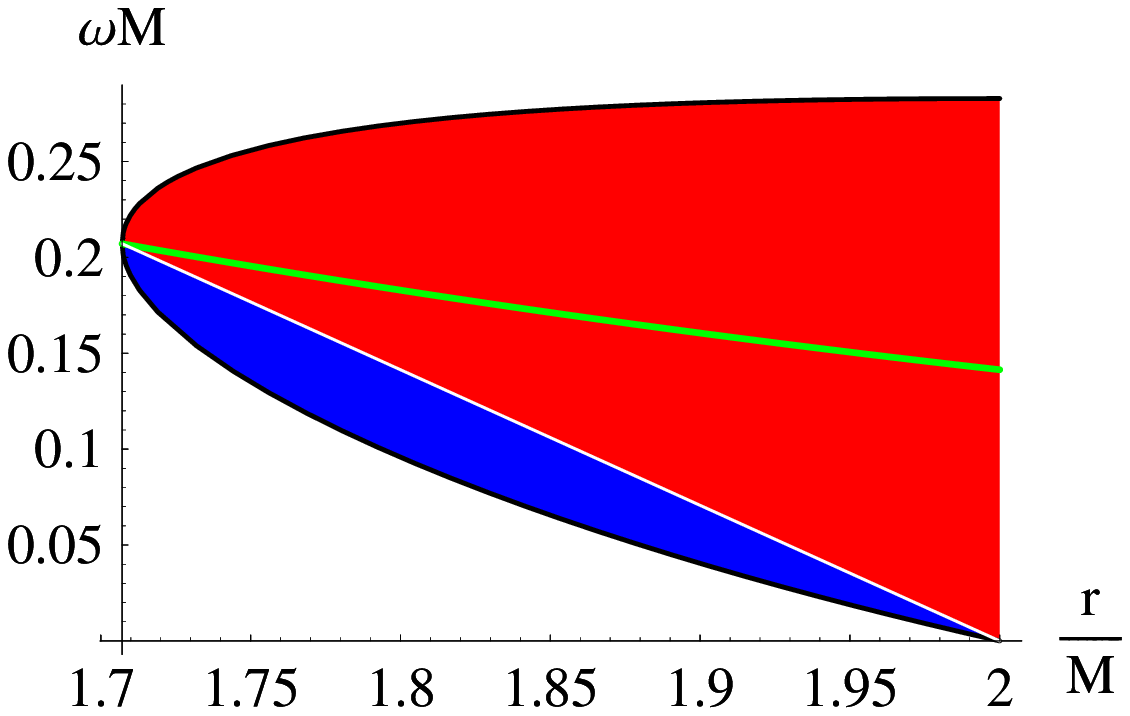} \\
\hspace{-7mm}(a) \hspace{55mm} (b) \\
\includegraphics[width=60mm]{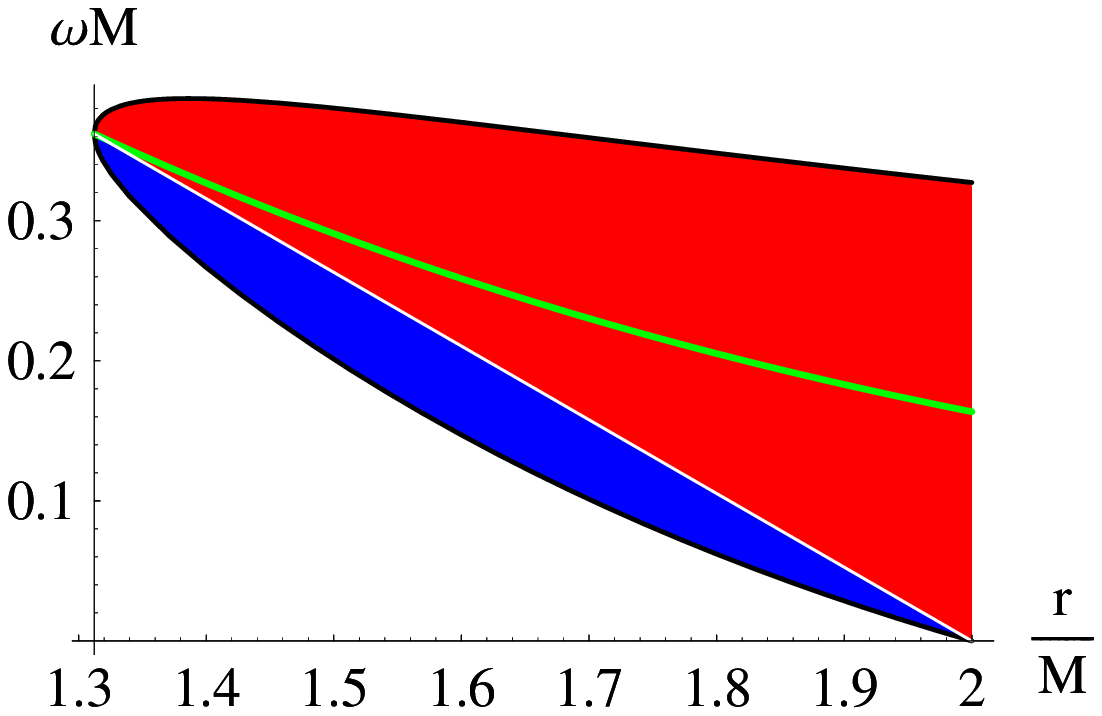} \ \ \ \
\includegraphics[width=63mm]{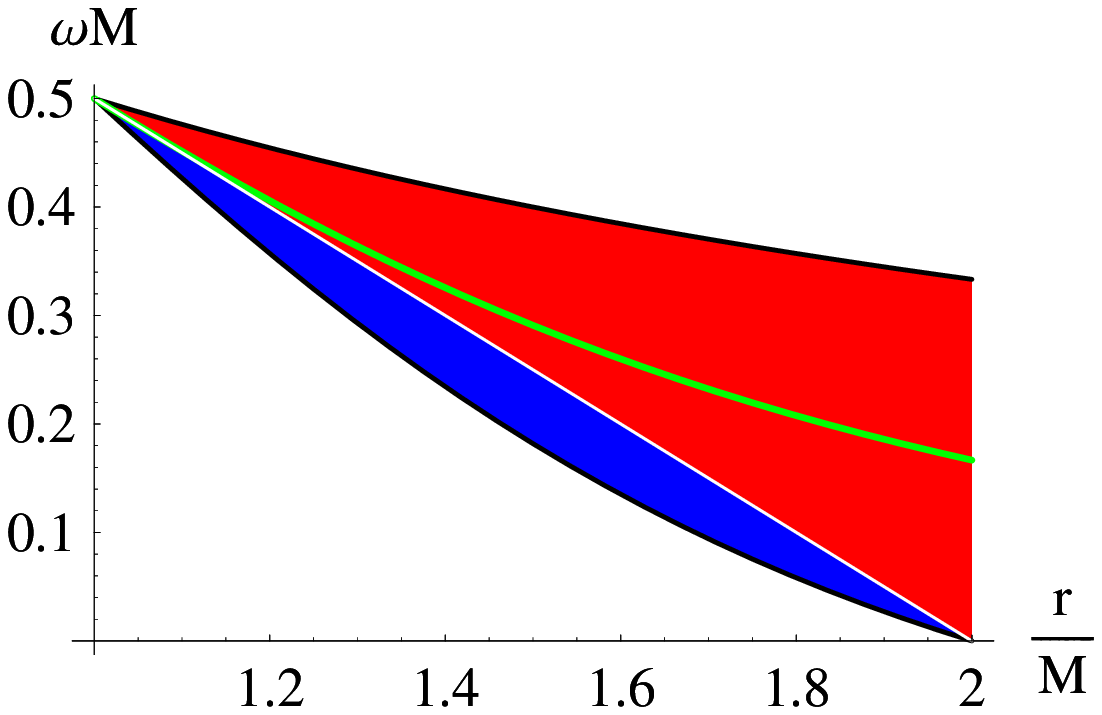}\\
\hspace{-7mm}(c) \hspace{55mm} (d) \\
\caption{Regions of possible angular velocities of particles in
ergosphere for $\theta = \pi/2$ and
a)~$a/M=0.5$, b) $a/M=1/\sqrt{2} $, c) $a/M=0.95$, d) $a/M=1$.
The red region --- angular velocities of particles with positive
energy, the blue region --- for particles with negative energy,
the white line --- $E=0$, the green line --- $J=0$.}
\label{FigOmeg}
\end{figure}

    Putting boundary values of expressions~(\ref{OmK3}), (\ref{OmK4})
for $m=0$ and $\Theta = 0$ into~(\ref{OmK2}) one obtains expressions
for $\Omega_1$ and $\Omega_2$ (see~(\ref{OmK1})).
    Taking into account that $d \varphi / d t $ due to~(\ref{OmK2})is growing
with the growth of $J/(EM)$ on each interval of continuity
$ \left( -\infty,\, \Sigma^2/(2 M^2 r a) \right)$
and $ \left( \Sigma^2/(2 M^2 r a),\, +\infty \right) $ one obtains
the division of the region of possible angular velocities in the ergosphere
on two regions: the lower for particles with negative energy and
the upper one for particles with positive energy as it is shown on
Fig.~\ref{FigOmeg}.
%%%%%%%%%%%%%%%%%%%%%%%%%%%%%%%%%%%%%%%%%%%%%%%%%

    Going in~(\ref{OmK2}) to the limit $ |J/(EM)| \to \infty $
one obtains the expression~(\ref{omE0}).
    That is why the frontier of regions for particles with positive and
negative energies is the line corresponding to the angular velocity
of particles with zero energy.
    In case of rotation in equatorial plane this line is the direct line.

    So the value of the angular velocity of the particle in the ergosphere
for the given value of the radial coordinate uniquely shows the sign of
the energy of the particle relative to the infinity and the value
of $J/(EM)$.
    Particles with zero energy are particles moving in ergosphere with angular
velocity~(\ref{omE0}), corresponding to the white line on Fig.~\ref{FigOmeg}.
    They divide by the value of angular velocity of rotation at given $r$
in the ergosphere particles with negative energies from particles with
positive relative to infinity energy.
    Particles with negative energy are particles rotating in
the ergosphere with the angular velocity less than~(\ref{omE0})!

    The light green line on Fig.~\ref{FigOmeg} corresponds to the angular
velocity of free falling particles with zero projection of angular momentum.
    Note that in this case the energy cannot be negative.

%%%% *****************************************************************
\section{The energy of collisions of particles in the centre of mass frame}
\label{secX}

    The energy in the centre of mass frame $E_{\rm c.m.}$ of two colliding
particles with rest masses~$m_1$ and $m_2$ is found by putting in square
the formula
    \begin{equation} \label{SCM}
\left( E_{\rm c.m.}, 0\,,0\,,0\, \right) = p^{\,i}_{(1)} + p^{\,i}_{(2)},
\end{equation}
    where $p^{\,i}_{(n)}$ --- 4-momenta of particles $(n=1,2)$.
    Due to $p^{\,i}_{(n)} p_{(n)i}= m_n^2$ one obtains
    \begin{equation} \label{SCM2af}
E_{\rm c.m.}^{\,2} = m_1^2 + m_2^2 + 2 p^{\,i}_{(1)} p_{(2)i} .
\end{equation}
    For free falling particles with energies $E_1$ and $E_2$
(relative to infinity) and angular momenta $J_1, J_2$
from equations of geodesics one obtains
    \begin{eqnarray}
E_{\rm c.m.}^{\,2} = m_1^2 + m_2^2 - \frac{2}{\rho^2 } \sigma_{1 \theta} \sigma_{2 \theta}
\sqrt{ \Theta_1 \Theta_2 } +  \hspace{21mm}
\nonumber \\
+\, \frac{2}{\Delta \rho^2} \left[ E_1 E_2 \Sigma^2 - 2 M r a (E_1 J_2 + E_2 J_1 ) -
J_1 J_2 \frac{S}{\sin^2 \! \theta} - \sigma_{1 r} \sigma_{2 r} \sqrt{ R_1 R_2}
\right].
\label{Col3}
\end{eqnarray}

    Let us find the  expression of the energy in the centre of mass frame
through the relative velocity~$ v_{\rm rel}$ of particles at the moment
of collision~\cite{BanadosHassanainSilkWest10}.
    In the reference frame of the first particle one has for
the components of 4-velocities  of particles at this moment
    \begin{equation} \label{Relsk01}
u_{(1)}^i = (1,0,0,0), \ \ \
u_{(2)}^i = \frac{(1, \, \mathbf{v}_{\rm rel}) }{\sqrt{1\!- v_{\rm rel}^2}}.
\end{equation}
    So \ $ u_{(1)}^i u_{(2) i} = 1 \left/ \sqrt{1- v_{\rm rel}^2}\right. $,
    \begin{equation} \label{Relsk02}
v_{\rm rel} = \sqrt{1- \left( u_{(1)}^i u_{(2) i} \right)^{-2}}\,.
\end{equation}
    These expressions evidently don't depend on the coordinate system.

    From~(\ref{SCM2af}) and~(\ref{Relsk02}) one obtains
    \begin{equation} \label{Relsk03}
E_{\rm c.m.}^{\,2} = m_1^2 + m_2^2 +
\frac{2 m_1 m_2}{\sqrt{1 \!- v_{\rm rel}^2}}
\end{equation}
    and the nonlimited growth of the collision energy in the centre of mass
frame occurs due to growth of the relative velocity to the velocity of
light~\cite{Zaslavskii11}.

    If particles move in radial coordinate in one direction
($\sigma_{1 r} \sigma_{2 r} =1$) one obtains in the limit $r \to r_H$ by
solving in the last term in~(\ref{Col3}) the indeterminacy of the form $0/0$,
for collisions on the horizon
    \begin{eqnarray}
E_{\rm c.m.}^{\,2} = m_1^2 + m_2^2
- \frac{2}{\rho^2 } \sigma_{1 \theta} \sigma_{2 \theta} \sqrt{ \Theta_{1 H}
\Theta_{2 H}} + \left( m_1^2 + \frac{\Theta_{1 H}}{\rho_H^2}
\right) \frac{J_{2 H} - J_2}{J_{1 H} - J_1 } +
\nonumber \\
+\, \left( m_2^2 + \frac{\Theta_{2 H}}{\rho_H^2} \right)
\frac{J_{1 H} - J_1}{J_{2 H} - J_2 } +
\frac{\rho^2_H} { 4 M^2 r_H^2 \sin^2 \! \theta } \frac{( J_{1 H} J_2 -
J_{2 H} J_1)^2}{(J_{1 H} - J_1 ) (J_{2 H} - J_2 )}
\label{Col4}
\end{eqnarray}
    (see also~\cite{HaradaKimura11}).
    If one of the particles has the angular momentum projection $J =J_H$
(the critical particle) and the other particle has $J \ne J_H$
then the energy of collisions is divergent on the horizon.
    For extremal rotating black holes this was found in the paper of
Ba\~{n}ados-Silk-West~\cite{BanadosSilkWest09} (BSW-effect).
    For nonextremal black holes in the vicinity of the horizon $J =J_H$
is not possible however in case of multiple collisions it is
possible~\cite{GribPavlov2010,GribPavlov2011} to have $J$
very close to $J_H$ for $r \to r_H$.

    Very large energy of collisions can be obtained if due to multiple
collisions or external field the particle has large in absolute value
but negative angular momentum
projection~\cite{GribPavlov2013,GribPavlov2012} for the fixed value of
the particle energies due to~(\ref{lHmdd}).
    As it was obtained in~\cite{GribPavlov2013} for the collision in
ergosphere one has
    \begin{equation}
E_{\rm c.m.}^{\,2} \approx J_2 \frac{r^2 -2 r M +a^2 \cos^2 \! \theta}
{\rho^2 \Delta \sin^2 \! \theta}
\left( \sigma_{1r} \sqrt{J_{1 +}- J_1} -
\sigma_{2r} \sqrt{J_{1 -}- J_1} \right)^2 ,  \ \ J_2 \to - \infty.
\label{KerrJBner}
\end{equation}
    In case $J \to -\infty$ for fixed energy and mass of the particle
the trajectory has the form shown in Fig.~\ref{Fig}.
    Unbounded growth of the energy of collision with growing $-J$
due to~(\ref{Relsk03}) is conditioned as for BSW effect by the growth of
the relative velocity of particles to the velocity of light.
    Differently from the BSW effect this effect can take place at
any point of the ergosphere.

    The energy of direct collisions ($\sigma_{1 r} \sigma_{2 r} =-1$)
is divergent on the horizon
    \begin{equation}
E_{\rm c.m.}^{\,2} \sim \frac{4 a^2 }{\Delta \rho^2}
(J_{1H} - J_1) (J_{2H} - J_2) \to \infty , \ \ \ r \to r_H,
\label{Col6}
\end{equation}
    if $J_i  \ne J_{i H}$.
    This way of getting ultrahigh energy is possible for particles one
of which moves along white hole geodesic~\cite{GribPavlov2015,GribPavlov2015b}.

    Formulas~(\ref{SCM})--(\ref{Col6}) are valid for any colliding particles
as with positive as with negative (relative to infinity) energy.
    For any value of the energies of particles the energy in the centre of
mass frame satisfies the inequality
    \begin{equation}
E_{\rm c.m.} \ge m_1 + m_2,
\label{Enm1m2}
\end{equation}
    because colliding particles in the centre of mass frame are moving one
to another with some velocities.
    All three ways of getting unboundedly high energy of collisions are also
possible for particles with negative (zero) energy.

    The physical importance of these resonances is due to possible conversion
of dark matter particles into visible ones in the ergosphere of
astrophysical black holes~\cite{GribPavlov2007AGN,GribPavlov2008KLGN}.

    The results of some calculations of the energy of collisions of two particles
with equal masses $m$ one falling from the infinity on the black hole
with the other particle with positive, zero and negative energy
are given in Ref.~\cite{GribPavlov2016}.

%%%% *****************************************************************
\section{Negative and zero energies in rotating coordinate frame}
\label{secVr}

    Following~\cite{LL_II}, \S\,89, introduce notations for inertial
cylindrical space coordinates $r'$, $\varphi'$, $z'$ and time $t$.
    The interval in this coordinate system is
    \begin{equation}    \label{v1}
d s^2 = c^2 dt^2 - d r^{\prime\,2} - r^{\prime\,2} d \varphi^{\prime\,2}
- d z^{\prime\,2}.
\end{equation}
    Here we use the usual units so that $c$ is the velocity of light.
    For rotating system let us note the cylindrical coordinates
as $r$, $\varphi$, $z$.
    Let the rotation axis $z=z'$.
    Then
    \begin{equation}    \label{v2}
r'=r, \ \ \ \ z'=z, \ \ \ \ \varphi' = \varphi + \Omega t,
\end{equation}
    where $\Omega$ is the angular velocity of rotation.
    Putting~(\ref{v2}) into~(\ref{v1}) one obtains the expression for
the interval in rotational coordinate system
    \begin{equation}    \label{v3}
d s^2 = (c^2 - \Omega^2 r^2)\, dt^2 - 2 \Omega r^2 d \varphi\, d t -
d r^{2} - r^{2} d \varphi^{2} - d z^{2}.
\end{equation}
    Here we use the words ``coordinate system'' different from
the ``coordinate frame'' because we want to use expression~(\ref{v3})
for the interval for any $r$!

    Let the energy and momentum in the inertial system be $E'$, $\bf p'$.
    Then the four energy momentum vector due to~\cite{LL_II}, \S\,9,
in Descartes coordinates is
    \begin{equation}    \label{v6}
p'^{\, i} = \left( \frac{E'}{c}, \ {\bf p'} \right), \ \ \ \
p'_{\, i} = \left( \frac{E'}{c}, \ {\bf - p'} \right).
\end{equation}
    Compute the energy-momentum in rotating coordinate system
$p^i = \left( \partial x^i / \partial x^{\prime\,k} \right)  p^{\prime\,k}$.
    Then
    \begin{equation}    \label{v10}
p^{i} = \left( \frac{E'}{c}, \ p^{\prime\,r}, \ p^{\prime\,\varphi}
- \Omega \frac{E'}{c^2}, \ p^{\prime\,z} \right),
\end{equation}
    and covariant components are
    \begin{equation}    \label{v10kk}
p_{i} = \left( \frac{E'- \Omega r^2 p^{\prime\,\varphi}}{c},
\ - p^{\prime\,r}, \ -r^2 p^{\prime\,\varphi}, \ - p^{\prime\,z} \right)
= \left( \frac{E}{c},
\ - p^{\prime\,r}, \ - L_z, \ - p^{\prime\,z} \right).
\end{equation}
    So in rotating coordinate system the energy $E$ and
component of the angular momentum $L_z$ are
    \begin{equation}    \label{v10v}
E = E' - \Omega L'_z,  \ \ \ \ L_z = L'_z = r^2 p^{\prime\,\varphi}.
\end{equation}
    Surely both these values are constants of motion.
    It is important that covariant components but not contravariant components
of the energy momentum vector occur to be constants of motion.
    Contravariant components of the momenumt are proportional to
the corresponding velocities.

    For particles with mass $m$ in cylindrical coordinates of
Minkowski space-time one obtains
    \begin{equation}
\frac{E'^2}{c^2} - (p'^r)^2 - \frac{1}{r^2} L'^2_z - (p'^z)^2  = m^2 c^2.
\label{vmts}
\end{equation}
    Then we obtain from~(\ref{v10v}) the possible limits for energy
of particle in rotating coordinate system for fixed energy $E'$
in nonrotating system and given value $r$:
    \begin{equation}
E' \left( 1 - \frac{\Omega r}{c} \sqrt{1 - \frac{m^2 c^4}{E'^2}}\right) \le
E \le E' \left( 1 + \frac{\Omega r}{c} \sqrt{1 - \frac{m^2 c^4}{E'^2}}\right).
\label{v19}
\end{equation}
    For $r< c/ \Omega$ the energy $E>0$.
    For $r = c/ \Omega$ the zero energy $E=0$ can be achieved.
    For $r > c/ \Omega$ the energy $E$ can accept arbitrary values,
positive, zero and negative.

    This picture is similar to the case of the Kerr metric, when
the covariant components of the energy-momentum momentum of particle are
    \begin{equation}
p_t = E , \ \ \ p_r = - \sigma_r \frac{\sqrt{R}}{\Delta} , \ \ \
p_\theta = - \sigma_\theta \sqrt{\Theta} , \ \ \ p_\varphi = - J .
\label{pkovi}
\end{equation}
    The negative values of the energy $E$ in both cases can be
only in the region of space-time where $g_{00} <0$
(see the problem 17.9 in the book~\cite{LPPT}).

    It is well known that for a particle in rotating coordinate frame similar
to the Boyer-Lindquist coordinates for Kerr metric there exists region where
particle cannot be at rest $r > c/ \Omega$ and the frontier $r = c/ \Omega$
is the static limit.
    For the case of the Earth rotating along its axis with a period of one day
this distance is $\approx 4.1 \cdot 10^9$\,km,
which is much farther than Uranus but closer than Neptune.

    From formula~(\ref{v19}) one sees that negative and zero energy $E$
trajectories are  possible in the region  $r > c/ \Omega$
external to the static limit in rotating coordinate frame.
    The main difference of the energy of the particle in Kerr metric and
in rotating coordinate frame is that in Kerr metric one defines all
conserved values relative to space infinity while in rotating coordinates
the observer is located at the centre where the projection of angular momentum
of any particle is zero so $E=E'$.

\vspace{0mm}
%%%% ****************************************************************
{\bf Acknowledgements.}\,
This work was supported by the Russian Foundation for Basic Research,
grant No. 15-02-06818-a
and by the Russian Government Program of Competitive
Growth of Kazan Federal University.

%\vspace{5mm}
%%%% ****************************************************************


\begin{thebibliography}{99}
\setcounter{enumiv}{0}
\itemsep=0mm

\bibitem{MTW}
C.\,W.~Misner, K.\,S.~Thorne, J.\,A.~Wheeler,
\textit{Gravitation} (Freeman, San Francisco, 1973)

\bibitem{Chandrasekhar}
S.~Chandrasekhar, \textit{The Mathematical Theory of Black Holes}
(Clarendon Press, Oxford University Press, New York, Oxford, 1983)

\bibitem{NovikovFrolov}
I.\,D. Novikov, V.\,P. Frolov,
\textit{Physics of Black Holes} [in Russian] (Nauka, Moscow, 1986);
\, V.\,P.~Frolov, I.\,D.~Novikov,
\textit{Black Hole Physics: Basic Concepts and New Developments}
(Kluwer Acad. Publ., Dordrecht, 1998)

\bibitem{GribPavlov2016}
A.\,A. Grib, Yu.\,V. Pavlov,
\textit{Black holes and particles with zero and negative energy},
\href{http://arxiv.org/abs/1601.02592}
{arXiv:1601.02592} % [gr-qc].  %% (1--13).

\bibitem{GribPavlovVert}
A.\,A. Grib, Yu.\,V. Pavlov, V.\,D. Vertogradov,
\href{http://dx.doi.org/10.1142/S0217732314501107}
{Mod. Phys. Lett. A \textbf{29}, 1450110} (2014)
\\ Geodesics with negative energy in the ergosphere of rotating black holes.

\bibitem{Kerr63}
R.\,P. Kerr,
\href{http://dx.doi.org/10.1103/PhysRevLett.11.237}
{Phys. Rev. Lett. \textbf{11}, 237--238} (1963)
\\ Gravitational field of a spinning mass as an example of algebraically
special metrics.

\bibitem{BoyerLindquist67}
R.\,H. Boyer, R.\,W. Lindquist,
\href{http://dx.doi.org/10.1063/1.1705193}
{J. Math. Phys. \textbf{8}, 265--281} (1967)
\\ Maximal analytic extension of the Kerr metric.

\bibitem{GribPavlov2013}
A.\,A. Grib, Yu.\,V. Pavlov,
\href{http://dx.doi.org/10.1209/0295-5075/101/20004}
{Europhys. Lett. \textbf{101}, 20004} (2013)
\\ On the energy of particle collisions in the ergosphere of the rotating
black holes.

\bibitem{GribPavlov2012}
A.\,A. Grib, Yu.\,V. Pavlov,
\href{http://dx.doi.org/10.4213/tmf8480}
{Teor. Matem. Fiz. \textbf{176}, 60--68} (2013)
[English transl.:
\href{http://dx.doi.org/10.1007/s11232-013-0075-4}
{Theor. Math.}
\href{http://dx.doi.org/10.1007/s11232-013-0075-4}
{Phys. \textbf{176}, 881--887} (2013)],
\ Collision energy of particles in the ergosphere of rotating
black holes.

\bibitem{Zaslavskii13c}
O.\,B. Zaslavskii,
\href{http://dx.doi.org/10.1142/S0217732313500375}
{Mod. Phys. Lett. A \textbf{28}, 1350037} (2013)
\\ Acceleration of particles as a universal property of ergosphere.

\bibitem{LL_II}
L.\,D.~Landau, E.\,M. Lifshitz,
\textit{The Classical Theory of Fields} (Pergamon Press, Oxford, 1983)

\bibitem{BanadosHassanainSilkWest10}
M. Banados, B. Hassanain, J. Silk and S.\,M. West,
\href{http://dx.doi.org/10.1103/PhysRevD.83.023004}
{Phys. Rev. D \textbf{83}, 023004} (2011)
\\ Emergent flux from particle collisions near a Kerr black hole.

\bibitem{Zaslavskii11}
O.\,B. Zaslavskii,
\href{http://dx.doi.org/10.1103/PhysRevD.84.024007}
{Phys. Rev. D \textbf{84}, 024007} (2011)
\\ Acceleration of particles by black holes: Kinematic explanation.

\bibitem{HaradaKimura11}
T. Harada, M. Kimura,
\href{http://dx.doi.org/10.1103/PhysRevD.83.084041}
{Phys. Rev.~D \textbf{83},  084041} (2011)
\\ Collision of two general geodesic particles around a Kerr black hole.

\bibitem{BanadosSilkWest09}
M. Ba\~{n}ados, J. Silk, S.\,M. West,
\href{http://dx.doi.org/10.1103/PhysRevLett.103.111102}
{Phys. Rev. Lett. \textbf{103}, 111102} (2009)
\\ Kerr black holes as particle accelerators to arbitrarily high energy.

\bibitem{GribPavlov2010}
A.\,A. Grib, Yu.\,V. Pavlov,
\href{http://dx.doi.org/10.1134/S0021364010150014}
{JETP Letters \textbf{92}, 125--129 (2010)}
\\ On the collisions between particles in the vicinity
of rotating black holes.

\bibitem{GribPavlov2011}
A.\,A. Grib, Yu.\,V. Pavlov,
\href{http://dx.doi.org/10.1016/j.astropartphys.2010.12.005}
{Astropart. Phys. \textbf{34}, 581--586} (2011)
\\ On particle collisions in the gravitational field of the Kerr black hole.

\bibitem{GribPavlov2015}
A.\,A. Grib, Yu.\,V. Pavlov,
\href{http://dx.doi.org/10.1134/S0202289315010065}
{Grav. Cosmol. \textbf{21}, 13--18} (2015)
\\ Are black holes totally black?

\bibitem{GribPavlov2015b}
A.\,A. Grib, Yu.\,V. Pavlov,
\href{http://dx.doi.org/10.4213/tmf8910}
{Teor. Matem. Fiz. \textbf{185}, 77--85} (2015)
\ [English transl.:
\href{http://dx.doi.org/10.1007/s11232-015-0351-6}
{Theor. Math. Phys. \textbf{185}, 1425--1432} (2015)],
\ High energy physics in the vicinity of rotating black holes.

\bibitem{GribPavlov2007AGN}
A.\,A. Grib, Yu.\,V. Pavlov,
\href{http://dx.doi.org/10.1142/S0217732308027072}
{Mod. Phys. Lett. A \textbf{23}, 1151--1159} (2008)
\\ Do active galactic nuclei convert dark matter into visible particles?

\bibitem{GribPavlov2008KLGN}
A.\,A. Grib, Yu.\,V. Pavlov,
\href{http://dx.doi.org/10.1134/S0202289309010125}
{Grav. Cosmol. \textbf{15}, 44--48} (2009)
\\ Active galactic nuclei and transformation of dark matter into visible matter.

\bibitem{LPPT}
A.\,P. Lightman, W.\,H. Press, R.\,H. Price, S.\,A. Teukolsky,
\textit{Problem Book in Relativity and Gravitation}
(Princeton University Press, New Jersey, 1975)

\end{thebibliography}
\end{document}